\documentclass[a4,aps,showpacs,twocolumn,superscriptaddress]{revtex4}
\usepackage{graphicx}
\usepackage{dcolumn}
\usepackage{amsmath}

\begin{document}

\title{All information-theoretically secure key establishing protocols are quantum protocols}

\begin{abstract}
\rule{0ex}{3ex}
\noindent
{\bf
A theorem is proved which states that no classical key generating protocol could ever be provably secure. Consequently, candidates for provably secure protocols must rely on some quantum effect. Theorem relies on the fact that BB84 Quantum key distribution protocol has been proven secure. }
\end{abstract}

\author{Mario Stip\v cevi\' c}
\email{Mario.Stipcevic@irb.hr}
\affiliation{\footnotesize Rudjer Bo\v{s}kovi\'{c} Institute,
         Bijeni\v cka 54, P.O.B. 180, HR-10002 Zagreb, Croatia}

\maketitle


\noindent
One of the central problems of cryptography is the problem of establishing a secret key by two parties which do not share any previous secret. This problem appears very often in practice, for example when an anonymous buyer wants to safely pass its credit card number to a web shop, or when a secure computer communication is to be established using ssh or https protocols, etc.
\\

As proved by Shannon on a purely logical basis, two parties, let's name them Alice and Bob, can not establish any secret by a public discussion based on a publicly known protocol \cite{Shannon49}. Nevertheless, in practice, Alice and Bob can still establish a useful common secret (key) and there are two general approaches to do that: computationally secure and unconditionally secure protocols. With computationally secure protocols such as the Diffie-Hellman protocol \cite{DH76} Alice and Bob can compute their key easily (quickly) while Eve, who has all the information to do the same, faces practically insurmountable but in principle solvable computational problem. All key establishing protocols in wider use today are of that type. Computationally secure protocols are based on NP complete problems but unfortunately no such problem has been proved to be hard, in the sense of the complexity theory. The DH protocol is "classic" in a sense that it operates with no reference to quantum mechanics. It is governed by a pure mathematics and consequently can be performed on a Turing machine. Alice and Bob, for example, could be two memory locations in an ordinary computer (or computers). When we use "Secure Shell" program to access some distant computer via a secured channel, this is exactly what Alice and Bob are.
\\

By modifying the Shannon's model of a crypto-system, information-theoretically secure (i.e. provable unconditionally secure) systems become possible \cite{Maurer99}. In such systems an adversary simply does not have the necessary information to calculate the key. To make that possible, two modifications are necessary. Firstly, instead of requiring that the key established by legitimate parties is statistically independent of the information which (at best) an adversary can obtain, we allow for an arbitrarily small correlation between the two. Secondly, we require that the adversary can not obtain exactly the same information as the legitimate users. Provability comes with a price tag: Alice and Bob now must interact (more than once) in order to establish the secret key. 

\section*{The non-existence theorem}

\noindent
Examples of information-theoretically secure protocols are: BB84 \cite{BBB+92}, B92 \cite{Ben92} and EPR \cite{Eke91}. In particular, security of BB84 has been proven in \cite{May98, SP00}. Neither of these protocols can be entirely performed on a Turing machine because they assume a quantum channel. Such a channel is something rather exclusive. Almost all of contemporary computing technology and communication infrastructure is classical. Furthermore, in the absence of practical quantum repeaters, quantum channel is the main cause of severely limited range of the QKD \cite{DBCZ99}.
It is therefore legitimate to ask whether there is any classical IT 
secure protocol. 
The answer to that is found in the Theorem 1. below, 
but first we need three definitions.
\\

{\bf Definition 1.}  {\em Classical communication channel} is a device which has one input and at least two outputs such that perfect copies of the information sent to the input appear at the outputs.
\\

{\bf Definition 2.} A protocol is said to be {\em classical} if it assumes only classical channels and can be performed on a Turing machine.
\\

{\bf Definition 3.} A key establishing protocol is said to be
{\em information-theoretically secure} if it allows two legitimate parties,
who do not share a secret initially, to establish with a high probability 
a common string (key)
about which an adversary has only partial knowledge 
limited by the upper bound
known to (but not necessarily under control of) the legitimate parties.
\\

For example the BB84 protocol followed by the 
information reconciliation \cite{BBB+92} forms
an information-theoretically secure protocol.
This definition is slightly weaker then the more common one which requires 
the legitimate parties to be able to
limit the adversaries knowledge about the 
key to an arbitrarily
small fraction of the key. 
However, the two are almost equivalent.
Namely, any protocol which fits the Definition 3 can be complemented with 
the privacy amplification \cite{BBBM95}
in order to satisfy the stronger definition, whereas
any protocol that satisfies the stronger definition 
trivially satisfies the weaker one.
We are now ready to prove the following theorem.
\\

{\bf Theorem 1.} {\em Two parties initially sharing no secret can not establish a common 
string about which an adversary has only partial knowledge
by a purely classical protocol.}
\\

Other, more colloquial formulation of the theorem might be: "All information-theoretically secure key establishing protocols (in both weak and strong sense) are quantum protocols.".
\\

\begin{figure}[h]
\centerline{\includegraphics[width=75 mm,angle=0]{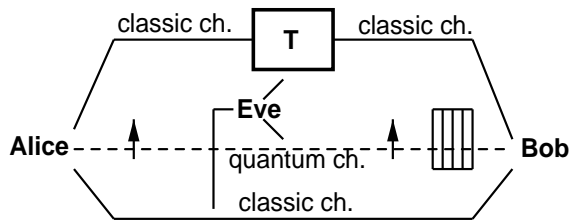}}
\caption{
          Alice and Bob can run BB84 protocol using one quantum and one classical channel. Alternatively they can run the same protocol after replacing the quantum channel with protocol T. (Setup for proving Theorem 1.)
}
\label{Fig1}
\end{figure}

{\bf Proof.} \
Imagine the situation as in Fig \ref{Fig1} where Alice and Bob are connected with one classical and one quantum channel so that they can perform BB84 protocol. Additionally, they are connected with channel(s) necessary to perform another information-theoretically secure protocol named T. We assume that T is a classical protocol and prove, by {\em reductio ad absurdum}, that this can not be true. 

In the setup in Fig \ref{Fig1}, Eve is familiar with both BB84 and T protocols and can make perfect copies of information going through all classical channels connected to Alice and Bob, and is free to make attempts to tamper with the quantum channel. In order to establish a common secret key Alice and Bob perform BB84 protocol using the quantum channel and the public channel (we assume BB84 protocol with polarized photons). Each of them has a local non-deterministic random bit generator. This protocol goes like this:
\\

\begin{enumerate}
\item Alice generates two bits in order to encode one of the 4 possible polarizations of photon whereas Bob generates one bit in order to choose one of the two polarizers;
\item Alice sends polarized photon to Bob;
\item At that moment Eve who is tampering with the quantum channel tries to find out which photon state has been emitted by Alice but she can not obtain full information because of no-cloning theorem;
\item Bob receives the photon and measures it with his polarizer. The result $r$ is interpreted as "0" or "1", according to the BB84 specification;
\item Bob sends to Alice the code of his polarizer, that is the bit he generated in the step 1, through the public channel;
\item If  Bob's polarizer matched photon polarization (meaning the measurement of the polarization was correct), Alice sends "OK" and the bit $r$ is kept by both Alice and Bob, otherwise it is discarded.
\item Alice and Bob repeat steps 1 through 6 any number of times to obtain each a string of desired length.
\item When the desired string length is reached, Alice anb Bob perform information reconciliation, which results in the final shared key.
\end{enumerate}

This protocol is information-theoretically secure. Its security relies entirely on the fact that Eve can not make a perfect copy of state emitted by Alice in the step 3 of each round, because no-cloning theorem \cite{no-cloning} of Quantum Mechanics applies to the quantum channel. However if the protocol T is secure then Alice and Bob can replace the quantum channel with T and perform a similar protocol in the following way:
\\

\begin{enumerate}
\item Alice generates two random bits, Bob generates one random bit; 
\item Alice, instead of generating a photon simply sends her two bits to Bob using classical protocol T;
\item At that moment Eve is listening to the protocol T but she can not obtain full information because T is information-theoretically secure;
\item Bob receives the bits from Alice, and calculates what he would measure in the case that Alice did send him a real photon and he used real polarizer according to his previously generated bit. The result $r$ is interpreted as "0" or "1", according to the BB84 specification. 
\item Bob sends to Alice the code of his "polarizer" that is the bit he generated in the step 1 through the public channel;
\item If  Bob's bit matched photon polarization code (meaning the measurement of the polarization would be correct), Alice sends "OK" and the bit $r$ is kept by both Alice and Bob, otherwise it is discarded.
\item Alice and Bob repeat steps 1 through 6 any number of times to obtain each a string of desired length.
\item When the desired string length is reached, Alice anb Bob perform information reconciliation, which results in the final shared key.
\end{enumerate}

This protocol also makes possible for Alice and Bob to obtain a partially secret key. But it differs from the first protocol only by the fact that the quantum channel in the step 3. has been replaced by a classical protocol T. This in turn means that the effect of quantum channel in BB84 can be achieved entirely in terms of the classical protocol T which can not be true because no-cloning theorem does not have a classical explanation, Q.E.D.
\\

Another way to prove the Theorem 1 would be to show that an unknown quantum state could be cloned with a help of a classical information-theoretically secure protocol.

\section{SKAPD protocol and the non-existence theorem}

It is the question of how the Theorem 1 relates to the SKAPD protocol of Maurer \cite{Mau93a} which is widely referred to being classical. 

\begin{figure}[h]
\centerline{\includegraphics[width=55 mm,angle=0]{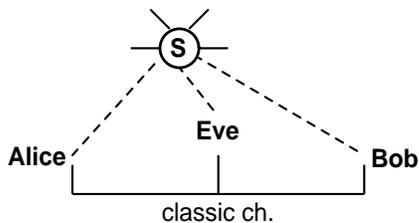}}
\caption{
          Maurer's "Secret Key Agreement by Public Discussion" (SKAPD) protocol
}
\label{Fig2}
\end{figure}

Maurer's protocol consists of three "phases" during which Alice and Bob (and possibly Eve) communicate over classical channels. Each phase (advantage distillation, information reconciliation and privacy amplification) is purely classical and can be performed on a Turing machine. Therefore it is tempting to think of it as being "classical". But, the three phases alone are not sufficient for establishing the key: they can not even begin without Alice and Bob (and Eve) having their initial binary strings. We are therefore coerced to conclude that there is another hidden but crucial "0-th" phase. In the Maurer's "satellite scenario" (Fig \ref{Fig2}) a satellite S emanates a weak signal of a random binary string, S$_0$. Alice, Bob and Eve are "connected" to the satellite by separate channels in order to obtain three different (but largely overlapping) copies of a binary string. Each of the parties uses its own copy to perform subsequent phases. Crucial requirement in the 0-th phase is that Eve can not copy (without errors) the information that is sent to either Alice or Bob. This means that, by definition, Alice and Bob are not connected to the satellite by classical channels. Further requirement is that the satellite sends a very weak signal such that Eve is not able to receive faithful copy of S$_0$ (in which case Alice and Bob have a zero chance of establishing the key). This requirement pushed to its extreme, by allowing Eve to have better and better receiving antenna, would eventually lead to the requirement that the satellite should only send one quantum of electromagnetic radiation at a time, thus apparently realizing quantum channels with Alice, Bob and Eve. One possible model of this low-power communication is that the satellite encodes zeros and ones with quantum states (polarizations) just the way Alice does in the BB84. But instead of sending only one photon per bit, it sends many photons of the same state (per bit) in all directions so that every participant would be able to receive multiple copies of the same state. Having received multiple copies corresponding to a given bit, parties would be able to measure the state but still not with absolute certainty. At the end, parties would obtain similar but not equal copies of the initial string S$_0$, which is exactly what is required.
Even more interesting is the fact that the security of this phase (in the present model) clearly comes from the quantum effects but not from the no-cloning theorem because no-cloning theorem applies only to single measurement of an unknown state whereas here one is allowed to perform multiple measurements.
\\

Scenarios other than the "satellite scenario" have also been worked out: for example digitizing of a publicly announced patch of the Moon's surface, or listening to a known weak radio source in deep Space, but no scenario is known in which zeroth phase is purely classical.
\\

To conclude, existence of the information-theoretically secure SKAPD protocol by itself does not contradict the Theorem 1.
\\

An additional support for the Theorem 1 comes from the recent work of \cite{tomographic} where striking equality has been found in noise thresholds for a family of quantum key-generating protocols and the SKAPD protocol indicating that these protocols might have more in common than previously realized.

\section*{Discussion}

\noindent
It could be argued that since we live in the quantum world all practical protocols are necessarily quantum. However, this is not what the Theorem 1 is about. 
Theorem 1 states that is is impossible to even {\em imagine} an information-theoretically secure protocol that would rely only on notion of classical channels and Turing machines. And since today communications and computers are to a very high degree classical, it means that they are unsuitable for information-theoretically secure key establishing protocols. 
Theorem 1 spells out a simple principle: information-theoretically  secure protocols must rely on some quantum effect, but not necessarily the no-cloning theorem.

\section*{Conclusion}

\noindent
By virtue of the Theorem 1 no classical key generating protocol could ever be proven secure. To illustrate this it has been shown that a provably secure protocol SKAPD, previously thought to be classical, is in fact not entirely classical. On the other hand, quantum protocols which have been proven secure so far are very impractical because of limited range, expensive gadgetry and generally non-existing infrastructure of quantum channels needed to support them on a wider scale. While some of these limitations could be overcame with further technological development, we hope that the Theorem 1 would inspire new ideas for more practical quantum protocols.

\end{document}